\documentclass[12pt]{aastex} 
\usepackage{emulateapj5}

\newcommand{\be}{\begin{equation}}
\newcommand{\ee}{\end{equation}}

\newcommand{\cm}{\,{\rm cm}}
\newcommand{\m}{\,{\rm m}}
\newcommand{\s}{\,{\rm s}}
\newcommand{\gm}{\,{\rm g}}
\newcommand{\km}{\,{\rm km}}
\def\refnew#1{(\ref{#1})} 
\newcommand{\au}{{\;\rm AU}}
\newcommand{\yr}{{\; \rm yr}}
\newcommand{\Myr}{{\; \rm Myr}}
\newcommand{\Gyr}{{\; \rm Gyr}}
\newcommand{\pomega}{\varpi}

\begin{document}

\title{Final Stages of Planet Formation} 

\author{Peter Goldreich$^{1,2}$, Yoram Lithwick$^{3}$, and Re'em Sari$^{2}$}
\affil{$^1$IAS, Princeton NJ, $^2$Caltech, Pasadena CA,
$^3$UCB, Berkeley CA \\
email: pmg@ias.edu, yoram@astro.berkeley.edu, sari@tapir.caltech.edu
}


\begin{abstract}

We address three questions regarding solar system planets. What
determined their number?  Why are their orbits nearly circular and
coplanar? How long did they take to form?  

Runaway accretion in a disk of small bodies resulted in a tiny
fraction of the bodies growing much larger than all the others. These
big bodies dominated the viscous stirring of all bodies. Dynamical
friction by small bodies cooled the random velocities of the big
ones. Random velocities of small bodies were cooled by mutual
collisions and/or gas drag.  Runaway accretion terminated when the
orbital separations of the big bodies became as wide as their feeding
zones. This was followed by oligarchic growth during which the big
bodies maintained similar masses and uniformly spaced semi-major
axes. As the oligarchs grew, their number density decreased, but their
surface mass density increased. We depart from standard treatments of
planet formation by assuming that as the big bodies got bigger, the
small ones got smaller as the result of undergoing a collisional
fragmentation cascade. It follows that oligarchy was a brief stage in
solar system evolution.

When the oligarchs' surface mass density matched that of the small
bodies, dynamical friction was no longer able to balance viscous
stirring, so their velocity dispersion increased to the extent that
their orbits crossed. This marked the end of oligarchy.  What happened
next differed in the inner and outer parts of the planetary system.
In the inner part, where the ratios of the escape velocities from the
surfaces of the planets to the escape velocities from their orbits are
smaller than unity, big bodies collided and coalesced after their
random velocities became comparable to their escape velocities. In the
outer part, where these ratios are larger than unity, the random
velocities of some of the big bodies continued to rise until they were
ejected. In both parts, the number density of the big bodies
eventually decreased to the extent that gravitational interactions
among them no longer produced large scale chaos. After that their
orbital eccentricities and inclinations were damped by dynamical
friction from the remaining small bodies.

The last and longest stage in planet formation was the clean up of
small bodies.  Our understanding of this stage is fraught with
uncertainty. The surviving protoplanets cleared wide gaps around their
orbits that inhibited their ability to accrete small
bodies. Nevertheless, in the inner planet system, all of the material
in the small bodies ended up inside planets. Small bodies in the outer
planet system probably could not have been accreted in the age of the
solar system. A second generation of planetesimals may have formed in
the disk of small bodies, either by collisional coagulation or
gravitational instability.  In the outer solar system, bodies of
kilometer size or larger would have had their random velocities
excited until their orbits crossed those of neighboring protoplanets.
Ultimately they would have either escaped from the sun or become
residents of the Oort cloud. An important distinction is that growth
of the inner planets continued through clean up, whereas assembly of
the outer planets was essentially complete by the end of oligarchy.
These conclusions imply that the surface density of the protoplanetary
disk was that of the minimum solar mass nebula in the inner planet
region but a few times larger in the outer planet region.  The
timescale through clean up was set by the accretion rate at the
geometrical cross section in the inner planet region, and by the
ejection rate at the gravitationally enhanced cross section in the
outer planet region. It was a few hundred million years in the former
and a few billion years in the latter. However, since Uranus and
Neptune acquired most of their mass by the end of oligarchy, they may
have formed before Earth!

A few implications of the above scenario are worth noting. Impacts
among protoplanets of comparable size were common in the inner planet
system but not in the outer. Ejections from the outer planet system
included several bodies with masses in excess of Earth after
oligarchy, and an adequate number of kilometer size bodies to populate
the Oort comet cloud during clean up. Except at the very end of clean
up, collisions prevented Uranus and Neptune from ejecting kilometer
size objects. Only Jupiter and, to a much lesser extent, Saturn were
capable of populating the Oort cloud with comets of kilometer size.

\end{abstract}

\section{Introduction}
 
Modern scenarios for planet formation may be broken down into several
stages.  The growth of the smallest gravitationally active bodies,
planetesimals, is mired in controversy \citep{Lis93,YS02}.  Orderly
growth by the merging of planetesimals is followed by runaway
accretion in which a small fraction of the bodies grow much larger
than all the others \citep{S72,WS89}.  When these big bodies are
sparse enough so that each dominates viscous stirring in its feeding
zone, runaway growth gives way to oligarchic growth during which the
big bodies grow in lockstep maintaining similar masses and uniformly
spaced orbits \citep{KI98}.  As oligarchs grow, their orbital spacing
increases and their number decreases. We investigate how oligarchy
ends and what happens after it does.  The plan of our paper is as
follows. We describe the conditions that pertain at the end of
oligarchy in \S \ref{s:before}. We show in \S \ref{s:after} that at
this stage dynamical friction from the small bodies is no longer able
to balance the mutual stirring of the big bodies. \S
\ref{s:completion} treats the regularization of the orbits of the big
bodies and the clean up of small bodies.  We summarize our findings in
\S \ref{ss:conclusions}.

A few definitions are in order. For simplicity, we consider two
classes of bodies, big ones and small ones, each composed of material
density $\rho$.  We denote the surface mass density, random velocity
dispersion, and radius of the former by $\Sigma$, $v$, and $R$, and of
the latter by $\sigma$, $u$, and $s$. The distance from and angular
velocity about the Sun are given by $a$ and $\Omega$. In our numerical
estimates, we set $\rho=5.5\gm\cm^{-3}$, the density of Earth, at
$1\au$ and $\rho=1.5\gm\cm^{-3}$, approximately the densities of
Uranus and Neptune, at $25\au$. For the condensible fraction of the
protoplanetary nebula, we adopt the surface densities $\sigma=
7\gm\cm^{-2}$ at $1\au$ and $\sigma=1.5\gm\cm^{-2}$ at $25\au$. The
former is just that appropriate to the minimum mass solar nebula
\citep{Hayashi81}, but the latter is enhanced sixfold relative to
it. This enhancement is designed to make the isolation mass in the
outer planet system comparable to the masses of Uranus and
Neptune. This is necessary, since the timescale for the accumulation
of the outer planets by coagulation of smaller isolation masses would
exceed the age of the solar system. The particular value of six
applies if half the mass had accreted into protoplanets by the end of
oligarchy. Instead, if most of it had, three would be the appropriate
enhancement factor.  In evaluating expressions containing the planet
mass, $M_p$, we use $1 M_\oplus$ for an inner planet and $15 M_\oplus$
for an outer one, where $M_\oplus\approx 6.0\times 10^{27}\gm$ is
Earth's mass.

It proves convenient to employ a
symbol $\alpha$ for the ratio between the radius of a body and that
of its Hill sphere, $R_H$:
\begin{equation}
\alpha\equiv \left(
{9\over 4\pi}{M_\odot\over \rho a^3}
\right)^{1/3}\approx
\cases{ 
234^{-1} & for $a = 1\au$ \cr
3800^{-1} & for $a =25\au$
} \, .
\end{equation}
Note that $\alpha$ is approximately the angle subtended by the sun at
$a$.  We also make use of the Hill velocity of the big bodies,
$v_H\sim \Omega R_H$; $v_H\sim \alpha^{1/2}v_{\rm esc}$, where $v_{\rm
esc}$ is the escape speed from the surface of a big body.

\section{At Isolation \label{s:before}}

Here we assume that oligarchy ends when approximately half the
original surface density has been accreted.\footnote{This assumption
is validated in \S \ref{s:after}} We refer to this as the epoch of
isolation.

The size, $s$, and velocity dispersion, $u$, of the small bodies are
uncertain.  They are also closely related: $u$ is set by an
equilibrium between viscous stirring by the big bodies and damping in
mutual collisions whose rate is inversely proportional to $s$.  We
assume that as the big bodies grow and viscous stirring intensifies,
the small bodies are collisionally fragmented. Fragmentation lowers
their random velocities making them easier to accrete. For simplicity,
in \S\ref{ss:ueqvH} we scale all quantities by the values they would
have if $u$ were fixed at the boundary between the shear and
dispersion dominated regimes, that is $u/v_H \sim 1$.  Generalization
to other values of $u/v_H$ is given in \S\ref{sizes} along with the
dependence of $u/v_H$ on $s$.

\subsection{Conditions at Isolation for $u\approx v_H$}
\label{ss:ueqvH}

With $u=v_H$ throughout oligarchy, $v\leq u$ with equality obtaining
at isolation (see eqns. [\ref{eq:vnusd}] \& [\ref{eq:vnudd}] in \S
\ref{ss:endoligarchy}).  Thus each oligarch accretes material from
within an annulus of half-width $2.5 R_H$ \citep{PH86}, so the
isolation radius
\begin{equation}
\label{eq:Riso}
R_{\rm iso}\approx \left({15\over 4}{\sigma
a\over\rho\alpha}\right)^{1/2} \approx \cases{1.3\times 10^3\km & for $a =
1\au$ \cr 2.3\times 10^4\km  & for $a = 25\au$ } \, .
\end{equation}
Equivalently, the isolation mass
\begin{equation}
\label{eq:Miso}
M_{\rm iso} \approx  {4 \pi \over 3} \rho R_{iso}^3 \approx
\cases{ 
8\times 10^{-3}M_\oplus 
& for $a = 1\au$ \cr
1.3\times 10^1  M_\oplus
& for $a = 25\au$
}\, .
\end{equation}
The number of oligarchs per unit logarithmic semimajor axis is
\begin{equation}
\label{eq:Niso}
N_{iso}=
{\pi\sigma a^2\over M_{\rm iso}}
\approx \\ \cases{100 & for $a = 1\au$ \cr 
9 & for $a = 25\au$ }\, .
\end{equation}
Equations \refnew{eq:Riso}-\refnew{eq:Niso} apply for $u<v_H$.  However,
from here to the end of \S \ref{ss:ueqvH} we specialize to $u=v_H$.

With $u=v_H$, the ratio of the accretion cross section to the
geometric one is $(v_{esc}/u)^2 \approx \alpha^{-1}$.
Thus the timescale from the start of oligarchic growth until isolation is 
\begin{equation}
\label{eq:tiso}
t_{\rm iso}\sim 
\Omega^{-1}\alpha^{1/2}\left(\rho a\over\sigma\right)^{1/2} \sim 
\cases{
10^5\yr & for $a=1\au$ \cr
10^7\yr & for $a=25\au$ }\ .
\end{equation}

At isolation, the escape velocity from the surface of an oligarch is given by
\begin{equation}
v_{\rm esc}=\left(2GM_{\rm iso}\over R_{\rm iso}\right)^{1/2}\approx 
\cases{2.3\km\s^{-1} & for $a=1\au$\cr
21\km\s^{-1} & for $a=25\au$}\, .
\label{eq:vesciso}
\end{equation} 
These values are to be compared with the escape velocity from solar
orbit, $44\km\s^{-1}$ at $1\au$ and $8.8\km\s^{-1}$ at $25\au$.

Viscous stirring of small bodies by oligarchs at isolation results in
collisions at speeds $u_{\rm col}\sim v_H$, so
\begin{equation}
u_{\rm col}\sim \alpha^{-1}R_{\rm iso}\Omega\approx \cases{
60\m\s^{-1} & for $a = 1\au$ \cr 140\m\s^{-1} & for $a = 25\au$ } \, .
\label{eq:vHiso}
\end{equation}

\subsection{Sizes and Velocities of Small Bodies 
\label{sizes}}

The collision rate is directly proportional to the size $s$ of the
small bodies.  To maintain $u/v_H\sim 1$ at isolation requires the
effective radius of the small bodies to take on the particular value
\begin{equation}
s_b\sim \alpha^{3/2}\left(\sigma a\over \rho\right)^{1/2}\sim
\cases{10\m  & for $a = 1\au$  \cr 1\m  &
for $a=25\au$ } \, .
\label{eq:sb}
\end{equation}

In the shear dominated regime, $s<s_b$, $u/v_H\sim s/s_b$. This does
not affect the $2.5\, R_H$ half width of an oligarch's feeding zone.
So $R_{\rm iso}$, $M_{\rm iso}$, and $N_{\rm iso}$ are still given by
equations \refnew{eq:Riso}, \refnew{eq:Miso}, and
\refnew{eq:Niso}. Moreover, $u_{\rm col}$ remains comparable to $v_H$
(eq. [\ref{eq:vHiso}]), the typical random velocity at which a small body
exits an oligarch's Hill sphere. However, as a consequence of the
reduced thickness of the disk of small bodies, $t_{\rm iso}\propto s$
provided $\alpha^{1/2}s_b<s<s_b$. There is no further reduction of
$t_{\rm iso}$ for $s<\alpha^{1/2}s_b$. We note that $t_{\rm iso}$ can
be remarkably small. For $s=\alpha^{1/2}s_b$,
\begin{equation}
t_{\rm iso}\sim \cases{10^4\yr & for $a=1\au$\cr 
10^5\yr & for $a=25\au$}  \, , 
\label{eq:mintiso}
\end{equation}
which follows from multiplying the values in equation \refnew{eq:tiso} by
$\alpha^{1/2}$.

In the dispersion dominated regime, $s>s_b$, $u/v_H\sim
(s/s_b)^{2/9}$.  Thus $u_{\rm col}\sim u$, and the width of an
oligarch's feeding zone $u/\Omega\sim (s/s_b)^{2/9}R_H$. Consequently,
to obtain values for $R_{\rm iso}$, $M_{\rm iso}$, $N_{\rm iso}$, and
$t_{\rm iso}$ appropriate to the dispersion dominated regime, we must
multiply those given in equations \refnew{eq:Riso}, \refnew{eq:Miso},
\refnew{eq:Niso}, and \refnew{eq:tiso} by factors of $(s/s_b)^{1/9}$,
$(s/s_b)^{1/3}$, $(s/s_b)^{-1/3}$, and $(s/s_b)^{5/9}$, respectively.

\subsection{Summary}

There are several messages to take away from this section. 
\begin{itemize}
\item A short isolation timescale requires accretion of small
bodies.
\item As the result of viscous stirring by oligarchs, small bodies
suffer collisions at velocities $\geq v_H$ that we assume are
sufficient to fragment them.
\item We note that $t_{\rm iso}$ as used by us measures the duration
of oligarchy. It is quite possible that it is shorter than the
timescales for orderly and runaway growth, the two stages that precede
it.
\item Models for Uranus and Neptune imply that each planet contains a
few $M_\oplus$ of hydrogen and helium \citep{Guillot99}. This is
consistent with a formation timescale $\sim 10^7\yr$ which requires
that these planets grew by accreting mainly meter size or smaller
bodies.
\item If collisional fragmentation continues to small enough sizes, the
disk of small bodies would be optically thick. Then it would be
described by fluid rather than by particle dynamics.
\item The isolation mass for the minimum mass solar nebula is smaller
than the planet mass, and by a much greater margin in the inner planet
system than in the outer. 
\end{itemize}
 
\section{Beyond Oligarchy}
\label{s:after}

\label{ss:endoligarchy}

We show below that oligarchy ends when $\Sigma\approx \sigma$. This
result applies to accretion in both shear and dispersion dominated
regimes. 

\subsection{Shear Dominated Accretion, $u<v_H$}

In this regime the big bodies heat each other and are cooled by
dynamical friction from the small bodies at the rates\footnote{The
heating rate given in the literature is proportional to $(v_H/v)^2$
instead of to $v_H/v$. \cite{GLS04} show that the latter is correct.}
\begin{equation}
{1\over v}{dv\over dt} \sim
{\Sigma \Omega \over \rho R} \alpha^{-2} {v_H \over v}
-{\sigma \Omega \over \rho R} \alpha^{-2} 
\label{eq:vsubhill}
\end{equation}
At equilibrium, $dv/dt=0$,
\begin{equation}
v\sim {\Sigma \over \sigma} v_H\, .
\label{eq:vnusd}
\end{equation}
This justifies the use of the heating rate appropriate to $v<v_H$ in
equation \refnew{eq:vsubhill}. 

\subsection{Dispersion Dominated Accretion, $u>v_H$}

It suffices to consider the regime in which $v>v_H$ as well as
$u>v_H$. Under these conditions the big bodies heat each other and are
cooled by dynamical friction from the small bodies at the rates
\begin{equation}
{1\over v}{dv\over dt}\sim
{\Sigma \Omega \over \rho R} \alpha^{-2}\left(v_H \over v\right)^4
-{\sigma \Omega \over \rho R} \alpha^{-2}\left(v_H\over u\right)^4\,  .
\label{eq:vsuperhill}
\end{equation}
Equilibrium occurs at
\begin{equation}
{v\over u}\sim \left(\Sigma\over\sigma\right)^{1/4}\, .
\label{eq:vnudd}
\end{equation}

\subsection{Instability Of Protoplanet's Velocity Dispersion}

As soon as $\Sigma>\sigma$, the velocity dispersion of the big bodies
destabilizes.  This occurs because the typical relative velocity
between a big and small body is $v>u$, so equations
(\ref{eq:vsubhill}) or (\ref{eq:vsuperhill}) are modified
to\footnote{Provided $v<v_{\rm esc}$.}
\begin{equation}
{1\over v}{dv\over dt}= {\left(\Sigma -\sigma \right)\Omega\over
\rho R}\alpha^{-2} \left( {v_H\over v} \right)^4.
\label{eq:vinstability}
\end{equation}
Thus when $\Sigma >\sigma$, big bodies are heated faster than they are
cooled.  This marks the end of oligarchy. As $v$ increases, heating
and cooling both slow down, but heating always dominates cooling.
Eventually the orbits of neighboring big bodies cross. 

Because it is based on approximate rates for viscous stirring and
dynamical friction, the criterion, $\Sigma\sim \sigma$, for the onset
of velocity instability is also approximate. Our choice of six times
the minimum mass solar nebula surface density in the outer planet
region is based on assuming that $\Sigma=\sigma$ at isolation. If
instead, at the onset of velocity instability, the oligarchs contained
most of the mass, the appropriate enhancement factor would be slightly
above three.  N-body simulations of oligarch dynamics with the
addition of accurate analytic expressions for dynamical friction can
resolve this issue.

The consequence of the instability in the velocity dispersion differs
according to which is larger, the escape velocity from the surfaces of
the planets that ultimately form or the escape velocity from their
orbits. The ratio of these two escape velocities is
\begin{equation}
{\cal R}\sim \cases{0.3 & for $a=1\au\,\, \& \,\, M_p=M_\oplus$ \cr
2.3 & for $a=25\au\,\, \& \,\, M_p=15 M_\oplus$}\, .
\label{eq:escratio}
\end{equation}    
 
Before we proceed to discuss these two cases, we stress an essential
point that is central to the outcome of each. N-body planet systems
can possess long term stability. This behavior lies outside the realm
that naive calculations of planetary interactions can describe.  We
propose that in both cases, ${\cal R}<1$ and ${\cal R}>1$, the system
of big bodies evolves such that the surviving planets have close to
the smallest spacings allowed by long-term stability.

\subsubsection{Inner Solar System, ${\cal R}\ll 1$: Coalescence}

In regions where ${\cal R}\ll 1$, the big bodies' velocity dispersion
increases until it becomes of order the escape velocity from their
surfaces. At this point they begin to collide and
coalesce. Coalescence slows as the number of big bodies decreases and
their individual masses increase. 

The timescale for the formation of planet size bodies with radius
$R_{\rm p}$ whose orbits are separated by of order $a$ is just
\begin{equation}
t_{\rm coag}\sim \left(\rho R_p\over \sigma \Omega\right)\sim 10^8\yr
\,\, {\rm at\,\,} a=1\au\,\, {\rm for} \,\, R_p=R_\oplus\, .
\label{eq:tcoag}
\end{equation}
At a separation of order $a$, mutual interactions no longer produce
chaotic perturbations. Indeed, detailed N-body simulations of terrestrial
planet formation by \cite{Cham01} produce stable systems on a timescale
similar to $t_{\rm coag}$. 

What happens to the small bodies while the big ones are colliding and
coalescing? A significant fraction collide with and are accreted by
big bodies.  Additional small bodies are created in grazing collisions
between big ones \citep{LR02}.  This ensures that a significant residual
population of small bodies persists until the end of coalescence.

\subsubsection{Outer Solar System, ${\cal R}\gg 1$: Ejection}

In regions where ${\cal R}\gg 1$, $v$ reaches the orbital speed
$\Omega a$.  Some fraction of the big bodies become detached from the
planetary system and either take up residence in the Oort cloud or
escape from the sun. This continues until mutual interactions among
the surviving big bodies are no longer capable of driving large scale
chaos.

We estimate the ejection timescale as
\begin{equation} 
t_{\rm eject}\sim {0.1\over\Omega}\left(M_\odot\over
M_p\right)^2\sim 10^{9}
{\yr}\,\, {\rm at\,\,} a=25\au \, .
\label{eq:tesc} 
\end{equation} 
\cite{SW84} and \cite{DWL+04} report similar timescales for the
ejection of test particles placed on orbits between Uranus and
Neptune, the former from a crude impulsive treatment of scattering and
the latter from N-body integrations. A shorter timescale might apply
if bodies were transferred to and then ejected by Jupiter and
Saturn. A quantitative estimate of the transfer rate may be obtained
from equation \refnew{eq:tcross}.

As the random velocity of a big body increases, the rate at which it
accretes small bodies declines. Thus a substantial surface density of
small bodies is likely to remain after most of the big bodies have
been ejected. In the following section we argue that most of the mass
in these small bodies eventually is either injected into the Oort cloud
or escapes from the sun.

\section{Completion}
\label{s:completion}
Here we consider processes that took place at sufficiently late
times, $>10^8\yr$ in the inner planet region and $>10^9\yr$ in the
outer, that it seems safe to ignore effects of gas drag.

\subsection{Gap Clearing}
\label{ss:gaps}

Gaps were not important prior to isolation because the radial
spacing of big bodies was only a few times larger than the widths of
their feeding zones. But after the protoplanets achieved large scale
orbital stability, their radial spacing was much larger than their
Hill radii and wide gaps would have formed around their orbits. 

Gap formation is driven by the torque per unit mass \citep{GT80}
\begin{equation}
T_p\sim {\rm sgn}(x)\left(M_p\over M_\odot\right)^2{\Omega^2 a^6\over
x^4}\,
\label{eq:PLT}
\end{equation}
that a protoplanet exerts on material at distance $x=a-a_p$ from its
orbit.\footnote{Equation \refnew{eq:PLT} is obtained by a radial
smoothing of the torque which has peaks at mean motion resonances.}
The gap width increases with time according to
\begin{eqnarray}
{|x|\over a}&\sim& \left(M_p\over M_\odot\right)^{2/5}\left(\Omega
t\right)^{1/5}
\nonumber \\
 &\sim&\cases{0.6 \ t_{\rm Gyr}^{1/5} & for
$a=1\au$\cr 0.6 \ t_{\rm Gyr}^{1/5} & for $a=25\au$}
\label{eq:xoft} \ ,
\end{eqnarray}
where $t_{\rm Gyr}=t/10^9\yr$, and we neglect the presence of other planets. 

Because the disk's viscosity arises from random motions excited by the
protoplanet, the width of the gap is independent of the collision
rate.  Gap edges are sharp or diffuse depending upon whether
collisions damp the amplitudes of epicyclic oscillations excited at
conjunctions before or after their phases decohere \citep{BGT89}.  The
former would allow accretion, albeit inhibited, while the latter would
shut it off altogether.  The appendix provides additional details
about the excitation of random motions and the profiles of gap edges.

\subsection{Orbit Regularization}
\label{ss:regularization}

Either coagulation or ejection is likely to end with the surviving big
bodies moving on orbits with eccentricities and inclinations of order
${\cal R}\sim 0.3$ in the inner planet system and of order unity in
the outer planet system. The former is seen in N-body simulations of
the formation of terrestrial planets from a few hundred big bodies,
with no small bodies present \citep{Cham01}.  Such orbits do not
resemble those of solar system planets.  In reality, dynamical friction
by the residual small bodies tends to circularize and flatten the
orbits of the surviving protoplanets. We can compare the rate at which
dynamical friction reduces $v$ to that at which big bodies grow by
accreting small ones. For $v\gtrsim v_{\rm esc}$ both rates are based
on physical collisions and are of the same order. However, for
$u<v<v_{\rm esc}$, the rate at which $v$ damps exceeds that at which $R$
grows by the factor $(v_{\rm esc}/v)^2$ for $v_H<v<v_{\rm esc}$,
$\alpha^{-1}(v/v_H)$ for $\alpha^{1/2}v_H<v<v_H$, and $\alpha^{-1/2}$
for $v<\alpha^{1/2}v_H$. These comparisons apply to a planet that
either cannot or has yet to open a gap around its orbit. 

Dynamical friction continues to act after gap opening. Angular
momentum and energy are transferred between the planet and the
disk of small particles by torques that the planet exerts at
Lindblad and corotation resonances. \cite{WH98,WH03} used the standard
torque formula \citep{GT80} and concluded that the most potent
contributions to the damping of eccentricity and inclination are due
to torques at apsidal and nodal resonances. They assessed their
contributions to be larger, by factors of $\sim \Omega/|\dot \pomega|$
and $\sim \Omega/|\dot\Omega_{\rm np}|$, than those from torques at
standard first order corotation and Lindblad resonances.\footnote{The
symbols $\dot\pomega$ and $\dot\Omega_{\rm np}$ denote apsidal and
nodal precession rates.}  However, this result comes with a number of
caveats, especially in applications to disks in which self-gravity
dominates pressure in the dispersion relation for apsidal and nodal
waves. \cite{WH98,WH03} assume that these waves are excited at apsidal
and nodal resonances, then propagate away and ultimately damp. They
further assume that the resonances lie farther from the planet than
the first wavelengths of the waves. But the main excitation of these
waves may occur off resonance at gap edges, and their long wavelengths
suggest that they may have more of a standing than a propagating wave
character \citep{GoS03}. Each of these features, and especially the
latter, is likely to reduce the rates of eccentricity and inclination
damping, but by amounts that are difficult to reliably estimate.

\subsection{Clean Up}

What was the fate of the residual small bodies that remained after the
protoplanets had settled onto stable orbits?  At the end of oligarchy,
small bodies and protoplanets contributed comparably to the overall
surface density.  But today the mass in small bodies is much less than
that in planets. The asteroid belt contains most of the mass not in
planets inside the orbit of Jupiter, but it totals $\lesssim
10^{-3}M_\oplus$. Our knowledge of small bodies in the outer planet
region is less complete, but observations of perihelion passages of
Halley's comet limit the mass of a disk at $a\gg 30\au$ to be
$\lesssim 10(a/100\au)^3 M_\oplus$ \citep{HMW68,Yeomans86,HQT91}.

Clean up was both the last and longest stage in solar system
evolution. It is ongoing in both the asteroid and Kuiper belts. The
Oort comet cloud was probably populated during this stage.  We outline
our thoughts on clean up below. They are speculations based on
interweaving theory and observation.

\subsubsection{direct accretion of small bodies}

Accretion of small bodies by protoplanets is the most obvious mechanism
for clean up. The rate at which a protoplanet gains mass by accreting
small bodies with $u\sim v_H$ from gap edges at $|x|\lesssim 2.5R_H$ is
\begin{equation}
{1\over M_p}{dM_p\over dt}\sim {\sigma_0\over\rho
R_p}\alpha^{-1}\left(5R_H\over \Delta a\right)^4\Omega\, ,
\label{eq:Mdotgap}
\end{equation}
where $\Delta a$ is the distance between neighboring planets,
$\sigma_0$ is the surface density of the small bodies far from the
gap, and we assume that the gap's surface density profile obeys
$\sigma\propto x^4$ (see eq.[\ref{eq:profilepower}]).  A more relevant
expression is that for $t_{\rm clean}\equiv \sigma_0\vert
d\sigma_0/dt\vert^{-1} \approx 2 \pi\sigma_0 a \Delta a(dM_p/dt)^{-1}
$;
\begin{eqnarray}
t_{\rm clean}&\sim & 2\times 10^{-2}\alpha^{-1}\left(M_\odot\over
M_p\right)^2 \left(\Delta a\over a\right)^5\Omega^{-1} \nonumber \\
&\sim & \left(\Delta a\over a\right)^5\times \cases{8\times 10^{10}\yr
\, & {\rm for} \, $a=1\au$\,  \cr 7\times 10^{11}\yr \, & {\rm for}\,
$a=25\au$\, .}
\label{eq:cleanaccrete}
\end{eqnarray}
In both the inner and outer solar system, the spacing between planets
is $\Delta a\sim a/3$, so $t_{\rm clean}\sim 300 \Myr$ for $1 \au$,
and $t_{\rm clean}\sim 3\Gyr$ for $25\au$.  The latter time is
uncomfortably long. It would be a factor $\alpha^{1/2}\sim 1/60$
smaller for $u\lesssim \alpha^{1/2}v_H$. However, this introduces a
new problem. Maintaining such a low velocity dispersion requires
frequent collisions and therefore substantial optical depth.  This may
lead to sharp gap edges and consequently the absence of accretion.
See the Appendix for more discussion of gap structure.

\subsubsection{second generation planetesimal formation}
\label{sec:gen2}

Towards the end of oligarchy, small bodies attain random speeds of
order $10^2\m\s^{-1}$ (eq. [\ref{eq:vHiso}]).  Collisions at such high
speeds fragment them to sizes much smaller than a kilometer. After
orbit regularization the protoplanets are spaced by many times their
Hill radii and viscous stirring of the intervening small bodies is
considerably weaker. An estimate for the rms random velocity of the
small bodies is given in equation \refnew{eq:uvss}. With our standard
parameters it yields
\begin{equation}
u_{\rm rms}\sim \left(\rho s\over\sigma \right)^{1/2}\left(a\over
|x|\right)^{3/2} \times \cases{ 0.1\m\s^{-1} \, & {\rm for} \,
$a=1\au$ \, \cr 0.3\m\s^{-1}\, & {\rm for} \, $a = 25 \au$ \, , }
\label{eq:urms} 
\end{equation}
where $|x|$ is radial distance from the protoplanet. Even these small
rms velocities are likely to be much larger than the mean random
velocities.  That is because the protoplanet's torque is concentrated
at discrete mean motion resonances, and the nonlinear disturbances it
raises damp locally \citep{GT78}. These strongly stirred regions near
resonances make the dominant contributions to $u_{\rm rms}$.  A
semi-quantitative discussion of this point is provided in the
appendix.

Do larger bodies, referred to here as planetesimals, form under the
conditions that prevail after orbit regularization? We are unable to
answer this question with confidence.  Instead, we critique the
difficulties faced by coagulation and gravitational instability, the
leading candidates for planetesimal formation.

\medskip

\centerline{\it by coagulation}

\smallskip

Without gravitational focusing, coagulation is a lengthy process. To
double its mass, a body would have to pass through the disk a minimum
of $\rho s/\sigma$ times. A potential problem is that a small body's
rms random velocity estimated from equation \refnew{eq:urms} is
greatly in excess of the escape velocity from its surface,
\begin{equation}
v_{\rm esc}\sim \alpha^{-3/2}\Omega s\sim \cases{0.7(s/\km)\m\s^{-1} \,  &
{\rm for}\, $a=1\au$\cr 0.4(s/\km)\m\s^{-1} \, &
{\rm for}\, $a=25\au$\, .}
\label{eq:vescs}
\end{equation}
This  would imply that collisions lead to disruption rather than to
coalescence.  Only bodies larger than
\begin{equation}
s_{\rm crit}\sim {M_p^2\over M_\odot M_d}{a^4\over|x|^3}\sim\cases{
5\times 10^2\km \, & {\rm for} \, $a=1\au$\cr 2\times 10^5\km \, & {\rm
for}\, $a=25\au$\, }
\label{eq:scrit}
\end{equation}
have $v_{\rm esc}>u_{\rm rms}$, where in the numerical evaluation, we
have set the disk mass, $M_d\sim \sigma a^2$, equal to the planet
mass, $M_p$, and $|x|=a$.

It might be argued that equation \refnew{eq:urms} does not apply, that
chaotic stirring would not occur far from a planet. This is certainly
true for perturbations from a single planet moving on a nearly
circular orbit. However, N-body calculations by several groups show
that stable orbits between planets are rare; even those initialized
with low eccentricities and inclinations invariably become planet orbit
crossers \citep{GD90,HW93,GNV+99}. Nevertheless, there are a couple of
reasons to wonder whether coagulation might still occur. None of the
N-body calculations investigated the stability of orbits with initial
random velocities as small as a few meters per second, and none of
them included the small amount of damping that passage though the
particle disk would cause.

\medskip

\centerline{\it by gravitational instability}

\smallskip

Gravitational instability is another possible mechanism for the
formation of second generation planetesimals. It has the virtue of
being very fast. However, it also faces a problem. The formation of
solid bodies by gravitational instability requires the particle disk
to be optically thick.  Observations of thermal infrared radiation
from solar type stars constrain the frequency of protoplanetary
systems with optically thick disks.
 
Suppose that the random velocity of the small bodies falls below the
limit for gravitational instability. That is,
\begin{eqnarray}
u\lesssim u_{\rm stab}&\sim& \pi G\sigma/\Omega 
\nonumber \\
 &\sim&
\cases{10\cm\s^{-1}\, & {\rm for\, } $a=1\au$\,
,\cr 1\m\s^{-1}\, & {\rm for\, } $a=25\au$ \, .}
\label{eq:stab}
\end{eqnarray}  
Gravitational instabilities convert potential energy into kinetic
energy of random motions. The development of nonlinear overdensities
requires this energy to be dissipated at the collapse rate $\sim
\Omega$. Otherwise the random velocity dispersion would be maintained
near the margin of stability, that is $u\sim u_{\rm stab}$
\citep{Gammie01}.  Inelastic collisions are the only option for
dissipating energy in a particle disk. For the collision rate to match
the collapse rate, the particle disk would have to be optically thick,
$\sigma/(\rho s)\gtrsim 1$.  An optically thick particle disk might 
result from a collisional fragmentation cascade.

The maximum size of a solid body that can form by collapse without 
angular momentum loss in a gravitationally unstable disk is
\begin{equation}
s_*\sim 
\alpha^{-3/2}{\sigma\over\rho}\sim
\cases{
{50\ \m {\rm  \ for}\ 
a=1\au.}\cr
{2 \ \km {\rm  \ for}\ 
a=25\au.}
}
\label{eq:sstar}
\end{equation}
Rapid damping of random velocities suggests that this is the size of
first bodies that will form by gravitational instability. Since the
escape velocity from their surfaces is $u_{\rm stab}$, mutual
interactions could maintain their random velocities at an adequate
level to stabilize the disk.

\subsubsection{Inner Solar System}

We assume that most of the mass contained in small bodies at the end
of coalescence ended up in planets and that only a small fraction fell
into the sun or was ejected by Jupiter. This assumption warrants
scrutiny, but that will not be done in this paper. 

The timescale for clean up by the accretion of small bodies, as given
in equation \refnew{eq:cleanaccrete}, could be comparable to or, for
$u<v_H$, even shorter than that for coagulation and orbit
regularization. However, this should not be taken to imply that second
order planetesimals did not form during clean up.

\subsubsection{Outer Solar System}

\medskip
 
\centerline{\it difficulties with accretion}

\smallskip

Accretion of the small bodies would be the simplest solution to clean
up. Estimates based on equation \refnew{eq:cleanaccrete}, which
assumes $u\approx v_H$, suggest that it would take a time comparable
to the age of the solar system for Uranus and Neptune to clean up the
region between them which has $\Delta a\sim a/3$, and far longer for
Neptune to clean up material from outside its orbit where the gap size
would be larger (see [\ref{eq:xoft}]).  Although for $u\lesssim
\alpha^{1/2}v_H$ the accretion rate would be a factor $\alpha^{-1/2}$
larger, it would require the disk of small bodies to maintain a
substantial optical depth. This might result in sharp gap edges and a
negligible accretion rate.  Given those uncertainties and our crude
estimates, we cannot exclude the possibility of accretion.

A more serious issue for our scenario concerns the amount of material
that might have been accreted after isolation. Could Uranus and
Neptune have acquired most of their mass during clean up?  Suppose the
initial surface density was only twice that of the minimum mass solar
nebula and that half remained in the form of small bodies at
isolation. Then the isolation mass would have been about one tenth the
mass of the outer planets. After a fraction of the big bodies were
ejected, dynamical friction from the small ones would have damped the
random velocities of the survivors. These would then have resumed
accreting small bodies and, once their masses had grown sufficiently,
their velocity dispersion would have again become unstable. This cycle
would have repeated until all the small bodies were
accreted.\footnote{For the sake of argument, we assume that the
accretion rate would have been fast enough for this to happen.} The
end result would not have been very different from that of our
preferred scenario. However, unless the original surface density
exceeded twice that of the minimum mass solar nebula, the repeated
ejections would have left too little mass to form planets as large as
Uranus and Neptune. Also, without the formation of a second
generation of planetesimals the connection to comets would be lost (see
below).

\medskip

\centerline{\it conditions for ejection}

\smallskip

Ejection is the alternative to accretion. Our story implies that up to
$\sim 100M_\oplus$ of small bodies was ejected in connection with the
formation of Uranus and Neptune. Such a large mass ejection aided by
Jupiter and Saturn would have been accompanied by a substantial
outward migration of Uranus and Neptune\citep{FIP84}. It might even
have moved them outside the orbits of most of the material from which
they formed \citep{LM03}.

To examine the conditions needed for ejection, we consider the fate of
a small body with radius $s$ embedded in a sea of bodies with radii
$\leq s$, and with total surface density $\sigma$.  It collides with a
total mass of order its own on a timescale
\begin{equation}
t_{\rm col}\sim {\rho {s}\over \Omega\sigma} \sim 2
\left({s\over 1 \km}\right) \Myr  \  {\rm\ \  for \ 25 \ AU} .
\end{equation}
By comparison, the timescale for a collisionless test particle placed
on a low eccentricity orbit midway between Uranus's and Neptune's
orbit to become an orbit crosser is 
\begin{equation}
t_{\rm cross}\sim \left( {M_\odot\over M_p} \right)^2 \left( {\Delta
a\over 2a} \right)^5\Omega^{-1} \sim 5\Myr   \ .
\label{eq:tcross}
\end{equation}
The above \"Opik-type estimate \citep{O76} agrees quite well with
results from N-body simulations \citep{GD90,HW93,GNV+99}.  Hence only
bodies with $s$ larger than
\begin{equation}
s_{\rm cross}\sim {\sigma\over  \rho}\left(M_\odot\over
M_p\right)^2\left(\Delta a\over 2 a\right)^5 \sim 2\km
\label{eq:scross}
\end{equation} 
could have become orbit crossers.  But ejection takes much longer than orbit
crossing; $t_{\rm eject}\sim 1\Gyr$ (eq. [\ref{eq:tesc}]).  Only
bodies larger than
\begin{equation}
s_{\rm eject}\sim 
0.1 {\sigma\over \rho}
\left(
{M_\odot\over M_p}
\right)^2\sim
{\rm 500 \ km} 
\label{eq:seject}
\end{equation} 
could have been ejected by Uranus and Neptune in the presence of a
disk of smaller bodies. However, for Jupiter, equations
\refnew{eq:tesc} and \refnew{eq:seject} yield $t_{\rm eject}\sim
10^5\yr$ and $s_{\rm eject}\sim 6 \km$.

The Oort cloud is a repository for kilometer size bodies that probably
formed in and were ejected from the outer planet region. Current
estimates of the cloud's mass lie in the range 1-10 $M_\oplus$
\citep{W96}, with the size of pristine comets, $1\km\lesssim s\lesssim
10\km$ being a major part of the uncertainty. Detailed numerical
calculations that follow the ejection of test particles from the outer
planet region show that a few percent end up in the Oort cloud
\citep{DWL+04}.  These, together with the observed flux of new comets,
are taken to imply that the outer planets ejected a few hundred
$M_\oplus$ of kilometer size bodies.  Some fraction may have
originated in the vicinity of Uranus and Neptune and been transferred
via Saturn to Jupiter which then ejected them.

Simplified treatments by \cite{SW84} and \cite{Fer97} as well as
N-body simulations by \cite{DWL+04} show that 50-80 percent of test
particles initially placed between Uranus and Neptune are, in fact,
ejected by Jupiter. But these investigations did not include
collisional damping whose importance was first recognized by
\cite{SW01}, and further investigated by 
\cite{CM03}.
When this is accounted for, it yields a stronger result:
Jupiter and, to a much lesser extent, Saturn were responsible for
ejecting almost all of the kilometer size bodies into the Oort cloud.
For this scenario to work, kilometer size bodies must have formed out
of the much smaller collisional debris that existed at the end of
oligarchy (see \S \ref{sec:gen2}).

\section{Discussion}
\label{s:discussion}

\subsection{Conclusions}
\label{ss:conclusions}

The scenario sketched in this letter addresses some of the basic
problems in planet formation.

\begin{itemize}

\item The number and orbital spacing of the planets resulted from  
an evolution toward stability against large scale chaotic perturbations.

\item After the cessation of chaotic perturbations, dynamical friction
by the residual small bodies damped the orbital eccentricities and
inclinations of the surviving protoplanets.

\item Accretion during oligarchy involved small bodies created by a
collisional fragmentation cascade. This stage probably lasted for 
less than $10^5\yr$ in the inner solar system and less than $10^7\yr$
in the outer solar system.

\item The timescale for establishing the final configuration of
planetary orbits was a few hundred million years for the inner planet
system and a few billion years in the outer planet system. It was set
by the accretion rate at the geometrical cross section in the former,
and by the ejection rate at the gravitationally enhanced cross section
in the latter.\footnote{The timescale in the outer solar system could
have been much shorter if all ejections were done by Jupiter and
Saturn.}

\item Clean up of small bodies is a complicated and poorly explored
stage of planet formation. Small bodies in the inner solar system were
incorporated into planets. Those in the outer solar system were
probably ejected by Jupiter and Saturn, but that requires a second
generation of planetesimal formation.

\end{itemize}

\subsection{Influence of Gas}
\label{ss:gasdrag}

We have neglected the influence of gas.  Observations of young stars
indicate that protostellar disks dissipate in a few million years
\citep{HLL01,Strom93}.  We show below that although the presence of
gas would alter some of our numerical results, it would not affect our
picture qualitatively.

Gas drag can provide significant damping for the random velocities of
small bodies in addition to that due to inelastic collisions. Relative
to collisions, it is most effective in damping the random velocities
of bodies that are smaller than the mean free path of the gas
molecules.  For these, its damping rate obeys the same expression as
that due to collisions, but with the surface density of the small
bodies replaced by that of the gas.  We can account for the effects of
gas drag by considering $s$ to be an effective size for the small
bodies that can be less than their true size. This is a minor point
for our story since, as we have emphasized, the true size of the small
bodies is highly uncertain. Moreover, our main concern is with the
stages of planet formation that follow velocity instability and these
probably continue after the gas is gone.

\cite{Raf03e} explored the fast accretion of protoplanetary cores in
the presence of gas. His investigation runs parallel to the early
phases of ours.  However, it terminates at the onset of velocity
instability, when $v\sim v_H$.

A potentially more significant effect of gas drag that was not
considered by \cite{Raf03e} is its role in damping the random
velocities of the oligarchs. \cite{Ward93} shows that the gas damping
rate can be obtained from the damping rate due to small bodies by
substituting the surface density of gas for that of the
small bodies and the sound speed of the gas, $c_s$, for the
random velocity of the small bodies, $u$.\footnote{We note that at
isolation $v_H<c_s<v_{\rm es}$. Moreover we are assuming that
$v<c_s$.} By stabilizing the oligarchs' random velocities, gas drag
could have enabled them to consume all of the small bodies.

In the inner solar system, it is possible---though highly
uncertain---that much of the gas survived until isolation. Then the
full velocity instability of the oligarchs would have been delayed
until the surface density of gas declined to match that contributed by
oligarchs. After that the oligarchs would have excited their random
velocities up to their escape speeds.  Although most of the small
bodies would have been accreted before this happened, plenty of new
ones created in glancing collisions could have damped the orbital
eccentricities and inclinations of the planets that finally
formed.\footnote{Since it took the inner planets more than $100\Myr$
to form (eqn. [\ref{eq:tcoag}]), gas is unlikely to have contributed
to regularizing their orbits.}

Outer solar system planets, Uranus and Neptune, are believed to have
collected only a few earth masses of nebular gas. So it is likely that
most of the gas had disappeared prior to isolation in the outer solar
system. 

Gas drag must have been more significant in the formation of Jupiter
and Saturn. One might worry that the orbital decay of small particles,
which are an integral part of our scenario, would have been too fast
for them to have been accreted. Particles with stopping time
comparable to their orbital time drift fastest. Their orbits decay on
a timescale $\Omega^{-1} (a\Omega/c_s)^2 \sim 10^3\yr$.  By damping
the random velocities of small bodies, gas drag can protect them from
undergoing destructive collisions. This may result in larger bodies,
for which the drift timescales are longer. Moreover, gas drag could
have made the isolation timescale in the Jupiter-Saturn region as
short as $\sim 10^4\yr$ (see equation [\ref{eq:mintiso}]).

\acknowledgements 

This research was supported in part by NSF grants AST-0098301 and PHY99-07949 and
NASA grant NAG5-12037. We thank Mike Brown, Eugene Chiang, Luke Dones,
Martin Duncan, Shigeru Ida, Roman Rafikov, and David Stevenson for
helpful advice.

\renewcommand{\theequation}{A-\arabic{equation}}
\setcounter{equation}{0}  
\section*{APPENDIX}  
\label{s:appendix}

In this appendix, we consider how a planet affects a disk of small
bodies in which it is embedded.  We assume throughout that the
collision time between small bodies is longer than the time it takes
their epicyclic phases to decohere.  This applies provided the optical
depth of the small bodies at a distance $\vert x \vert >R_H$ from the
planet is less than $\tau_{\rm crit}\sim(R_H/\vert x\vert)^3$.  If
this condition does not hold, i.e., if $\tau>\tau_{\rm crit}$, then
the effective viscosity would be negative, and the planet would open a
gap with sharp edges \citep{BGT89}.

\bigskip

\centerline{\it Stirring}

\medskip

To determine the mean random kinetic energy of the small bodies, we
balance their energy loss rate in inelastic collisions, $(1-\epsilon^2)
u^2/t_{\rm col}$, against the sum of the rates of energy gain from
direct forcing by the planet, $(\Omega_p-\Omega)T_p$ \citep{BGT82},
plus viscous dissipation acting on the Keplerian shear,
$\nu(ad\Omega/da)^2$. The symbol $\epsilon<1$ denotes the coefficient
of restitution, assumed to be a decreasing function of impact
velocity, $\Omega_p$ is the orbital frequency of the planet, and
$T_p$ is the torque per unit mass exerted by the planet.  For
$|x|\ll a$ we obtain
\begin{equation}
\left(1-\epsilon^2\right){u^2\over t_{\rm
col}}\approx {3x\over 2a}\Omega T_p +
{9\over 4}\nu\Omega^2\, ,  
\label{eq:dErandt}
\end{equation}
where the kinematic viscosity
\begin{equation}
\nu\sim {u^2\over t_{\rm coll}}\Omega^{-2}
\label{eq:nu}
\end{equation}
\citep{GT78},\footnote{ We consider only the case $\Omega t_{\rm
col}>1$; more generally, \cite{GT78} show $\nu\sim u^2 t_{\rm
col}\left[1+(\Omega t_{\rm col})^2\right]^{-1}$ for circular,
Keplerian rotation.} and 
\begin{equation}
t_{\rm col}^{-1}\sim {\sigma \Omega\over\rho s}\approx \tau\Omega \, .
\label{eq:tcol}
\end{equation}
In the limit $T_p=0$, energy released from the Keplerian shear would
maintain $u$ at some small value set by the velocity dependence of
$\epsilon$. The velocity dispersion in unperturbed parts of Saturn's
rings, $u\lesssim 1\cm\s^{-1}$, is a practical example. Our interest
is in circumstances under which forcing by a planet results in an
equilibrium value of $u$ that is much larger than that produced solely
by viscous dissipation acting on the Keplerian shear. Under such
conditions
\begin{equation}
u\sim {M_p\over M_\odot}\left(\rho
s\over\sigma\right)^{1/2}\left(a\over |x|\right)^{3/2}\Omega a
\, .
\label{eq:uvss}
\end{equation}

To this point we have proceeded as though all small bodies were of the
same size. However, equation \refnew{eq:uvss} applies more generally
and yields the size dependence of the rms random velocity of a small
body subject to two limits. It must be larger than those which contain
most of the mass but small enough so that gravitational focusing does
not enhance its interactions with them.  In this generalized
interpretation, $\sigma$ must be interpreted as the total surface
density in small bodies of all sizes.

\bigskip

\centerline{\it Resonances}

\medskip

The planet's torque is concentrated at resonances. Equation
\refnew{eq:uvss} is a spatial average of the torques at mean motion
resonances. These dominate the heating of the random velocities of
small bodies. As the rms random velocity given by equation
\refnew{eq:uvss} is a spatial average, it is of limited
utility. 

A more complete picture is obtained by investigating the disturbances
raised by torques at individual principal mean motion resonances
\citep{GT80}.  Our starting point is the WKB dispersion relation for
non-axisymmetric waves of angular degree $m$ in a cold,
self-gravitating disk;
\begin{equation}
(\omega-m\Omega)^2=\kappa^2-2\pi G \sigma  |k| \, .
\end{equation}
For principal mean motion resonances, $\omega=m\Omega_p$, where the
pattern speed, $\Omega_p$, is equal to the planet's mean orbital
angular velocity.  At the Lindblad resonance, $|k|$=0 and
$|x|/a\approx 2/(3m)$.

Density waves are excited at each mean motion resonance. Their
properties have been extensively studied; we merely quote a few
relevant results \citep{GT78}. These are specialized to the case of a
near Keplerian disk for which $\kappa\approx \Omega$. A wave
propagates away from the resonance and the planet 
at the group speed
\begin{equation}
v_g={\pi G \sigma \over \Omega}\, .
\end{equation}
Its first wavelength
\begin{equation}
{\lambda_1 \over |x|}=\left({\sigma a^2 \over M_\odot}{a\over |x|}
\right)^{1/2}\, .
\end{equation}

At each encounter with the protoplanet, a disk particle receives a
kick sufficient to change its orbital eccentricity by
\begin{equation}
\Delta e\sim {M_p\over M_\odot}\left(a\over |x|\right)^2\, .
\label{eq:Dele}
\end{equation}
At a Lindblad resonance, successive increments in $e$ sum coherently
over a time comparable to that during which a disturbance propagating
at the group velocity crosses the first wavelength. The number of
encounters that occur in this time is
\begin{equation}
N\sim {\lambda_1\Omega\over v_g}{|x|\over a}\, ,
\label{eq:N}
\end{equation}
so that at resonance,
\begin{equation}
e_{\rm res}\sim N\Delta e\sim \left({M_p^2\over M_\odot \sigma
a^2}{a\over|x|}\right)^{1/2}\, .
\label{eq:eres}
\end{equation}
The nonlinearity of the wave, $\Delta\sigma/\sigma$, is of order the
ratio of the coherent epicyclic excursions to the wavelength. Near
resonance this gives
\begin{equation}
{a\over\lambda_1} e_{\rm res}\sim {M_p\over \sigma a^2}{a\over |x|}\, .
\label{eq:NL}
\end{equation} 

In our scenario, the disk and planet have comparable masses, and both
are much smaller than the solar mass. Thus the density waves reach
order unit nonlinearity within their first wavelengths and their first
wavelengths are much smaller than the distance between neighboring
resonances. Therefore, the waves only propagate a small fraction of
the distance between resonances before damping. All of the planet's
excitation of random velocities is concentrated in these narrow
regions where they damp.

\bigskip

\centerline{\it Gaps}

\medskip

A protoplanet clears a gap in the disk of small bodies in which it is
embedded. Epicyclic motions are excited when small bodies pass
conjunction with the protoplanet. Provided their phases decohere before
collisions damp their amplitudes, the gap edges will be diffuse rather than
sharp. Since the accretion rate in the shear dominated limit is
proportional to the surface density at $|x|\approx 2.5R_H$, it is
important to determine the gap's surface density profile. In order to
do so we must estimate the order unity coefficient
relating the kinematic viscosity to the rate at which energy per unit
mass is dissipated by inelastic collisions. We define $b$ through
\begin{equation}
\nu\Omega^2={4b\over 9}{u^2\over t_{\rm col}}\, .
\label{eq:nub}
\end{equation}
Combining equations \refnew{eq:dErandt} and \refnew{eq:nub}, we find
\begin{equation}
b=1-\epsilon_*^2\, ,
\label{eq:detb}
\end{equation}
where $\epsilon_*$ is the value of $\epsilon$ at which the equilibrium
velocity dispersion is obtained for stirring by the Keplerian shear in
the absence of a protoplanet.  Since $T_p=K/x^4$ (eq.[\ref{eq:PLT}]),
equation \refnew{eq:dErandt} also implies that 
\begin{equation}
{u^2\over t_{\rm col}}\approx {3\Omega K\over
2a(\epsilon^2_*-\epsilon^2)|x|^3}\, ,
\label{eq:u2tc}
\end{equation}

In steady-state, the torque per unit mass that the protoplanet exerts
on a small body, $K/x^4$, is balanced by the viscous torque. Thus
\begin{equation}
{K\over x^4}={{3\Omega a}\over 2\sigma }{d(\nu\sigma)\over dx} \ ,
\label{eq:torbal}
\end{equation}
where we neglect gradients of $\Omega$ and $a$, since they are much
smaller than those of $\nu\sigma$ for $|x|\ll a$.  Combining
equations \refnew{eq:nub}, \refnew{eq:detb},  \refnew{eq:u2tc}, and
\refnew{eq:torbal} gives
\begin{equation}
{d\over
  dx}\left({(1-\epsilon_*^2)\over(\epsilon_*^2-\epsilon^2)}{\sigma\over
  |x|^3}\right)={\sigma\over x^4}\, .
\label{eq:torbalpr}
\end{equation}
For circumstances where stirring by the protoplanet is much greater than
that due to the Keplerian shear, it is likely that $\epsilon\ll \epsilon_*$.
In this case equation \refnew{eq:torbalpr} yields the gap profile
\begin{equation}
\sigma\propto |x|^q,\,\, {\rm with}\,\,
q=3+\epsilon_*^2/(1-\epsilon_*^2)\, .
\label{eq:sigprofile}
\end{equation}

It is difficult to obtain a reliable estimate for
$\epsilon_*$. \cite{GT78} obtained an approximate solution of the
collisional Boltzmann equation for a model in which the particles were
represented by smooth spheres separated by many times their diameters.
They found $\epsilon_*\approx 0.63$, which implies
\begin{equation} 
\sigma\propto
\vert x \vert^{3.66}\, .
\label{eq:profilepower}
\end{equation}
For the purposes of the present paper, we are content to approximate this as
$\sigma\propto |x|^4$ .

\vfill
\eject

\bibliographystyle{apj}
\bibliography{beyond}

\end{document}